# The skew-$t$ factor analysis model

Tsung-I Lin*, Pal H. Wu, Geoffrey J. McLachlan, Sharon X. Lee

**Abstract**

Factor analysis is a classical data reduction technique that seeks a potentially lower number of unobserved variables that can account for the correlations among the observed variables. This paper presents an extension of the factor analysis model by assuming jointly a restricted version of multivariate skew $t$ distribution for the latent factors and unobservable errors, called the skew-$t$ factor analysis model. The proposed model shows robustness to violations of normality assumptions of the underlying latent factors and provides flexibility in capturing extra skewness as well as heavier tails of the observed data. A computationally feasible ECM algorithm is developed for computing maximum likelihood estimates of the parameters. The usefulness of the proposed methodology is illustrated by a real-life example and results also demonstrates its better performance over various existing methods.

## 1 Introduction

Factor analysis (FA), which originated from the work of Spearman (1904), is concerned with a way of summarizing the variability between a number of correlated variables; see, for example, Lawley and Maxwell (1971). The correlations between the variables under consideration are explained by their linear dependence on a usually much smaller number of unobservable (latent) factors. In particular, FA can be considered as an extension of principal component analysis (PCA), both of which are widely used statistical tools for reducing dimensionality by constructing linear combinations of the variables. Unlike PCA, which forms only a set of linearly uncorrelated representations explaining the most variance of the variables, FA seeks the most correlation among the variables by including additive independent errors on the observed variables. FA can also be viewed as a clustering method where the variables are described by the same factors which are grouped together.

*Tsung-I Lin · P. H. Wu,
Institute of Statistics, National Chung Hsing University, Taichung 402, Taiwan;
e-mail: tilin@nchu.edu.tw
Geoffrey J. McLachlan · Sharon X. Lee
Department of Mathematics, University of Queensland, St Lucia, 4072, Australia
e-mail: g.mclachlan@uq.edu.au

FA has been applied successfully to numerous problems that arise naturally in many areas, see Basilevsky (2008) for a literature survey. In the FA framework, errors and factors are routinely assumed to have a Gaussian distribution because of their mathematical and computational tractability. However, the traditional FA approach has often been criticized for the lack of stability and robustness against non-normal characteristics such as skewness and heavy tails. Statistical methods which ignore the departure of normality may cause biased or misleading inference. To remedy this weakness, authors such as McLachlan et al. (2007), Wang and Lin (2013) and Zhang et al. (2013) considered the use of the multivariate $t$ (MVT) distribution for robust estimation of FA models, known as the $t$FA model. To our knowledge, there is little extended work on simultaneously accounting for asymmetry and heavy-tailedness in such models.

When the data have longer than normal tails or contain atypical observations (the so-called outliers), the multivariate $t$-distribution has been shown to be a natural extension of the normal for making robust statistical inference (Lange et al., 1989; Kotz and Nadarajah, 2004) as it has an extra tuning parameter, the degrees of freedom (df), to regulate the thickness of tails. In many biological applications (e.g., Pyne et al., 2009; Rossin et al., 2011; Ho et al. 2012) and other applied prob-lems, however, the data often involve observations whose distributions are highly asymmetric as well as having fat tails. One way is to use mixtures of factor analyzers (MFA) as proposed by Ghahramani and Hinton (1997). This approach was developed further in McLachlan and Peel (2000; 2007) and McLachlan et al. (2003; 2007). To further reduce the number of free parameters to be estimated, Baek et al. (2010) and Baek and McLachlan (2011) introduced MFA with common component factor loadings (MCFA) before rotation of the factors to be white noise.

Over the past two decades, there has been a growing interest in proposing more flexible parametric families that can accommodate skewness and other non-normal features. In particular, the family of multivariate skew $t$ (MST) distributions (Azzalini and Capitaino, 2003; Jones and Faddy, 2003; Sahu et al., 2003; Azzalini and Genton, 2008) had receive recent attention. This family contain additional skewness parameters for modeling asymmetry and includes the MVT family as a special case.

This paper presents a robust extension of FA model by replacing the normality assumption for the latent factor and errors with the restricted multivariate skew $t$ (rMST) distribution, hereafter referred to as the skew $t$ factor analysis (STFA) model. The rMST distribution is reduced to the restricted multivariate skew normal (rMSN) distribution when the df approaches infinity. Both of which were originally used in Pyne et al. (2009) based on a restricted variant of the skew-elliptical distributions of Sahu et al. (2003). A comprehensive overview of their characterizations together with their conditioning-type and convolution-type representations can be found in Lee and McLachlan (2013a; 2013b). The proposed STFA model is a good alternative to deal with dimensionality reduction of multivariate data that have fat tails with strong degrees of asymmetry. The STFA includes the classical FA and $t$FA models as special cases and thus would be more widely applicable.

The remainder of this paper is organized as follows. In Section 2, we establish the notation and briefly outline some preliminary properties of the rMSN and rMST distributions. Section 3 discusses the specification of STFA model and presents the development of ECM algorithm for obtaining the ML estimates of model parameters. In Section 4, we describe two simple ways of computing the standard errors of STFA model parameters based on the information-based method and the parametric bootstrap procedure. In Section 5, we illustrate the usefulness of the proposed method with a real-life data set. Some concluding remarks are given in Section 6 and technical derivations are sketched in Supplementary Appendices.

## 2 Preliminaries

We begin with a brief review of the restricted version of the MSN and the MST distributions and a study of some essential properties. To establish notation, we let $\phi_p(\cdot;\mu,\Sigma)$ be the probability density function of $N_p(\mu,\Sigma)$ (a $p$-variate multivariate normal distribution with mean $\mu$ and covariance matrix $\Sigma$); $\Phi(\cdot)$ be the cumulative distribution function (cdf) of the standard normal distribution; $t_p(\cdot;\mu,\Sigma,\nu)$ be the pdf of $t_p(\cdot;\mu,\Sigma,\nu)$ (a $p$-variate MVT with location $\mu$ and scale covariance matrix $\Sigma$ and degrees of freedom $\nu$); $T(\cdot;\nu)$ be the cdf of the Student's $t$ distribution with df $\nu$; $TN(\mu,\sigma^2;(a,b))$ be the truncated normal distribution for $N(\mu,\sigma^2)$ lying within a truncated interval $(a,b)$; $M^{1/2}$ denote the square root of a symmetric matrix $M$; $1_p$ denote a $p\times 1$ vector of ones; $I_p$ be the $p\times p$ identity matrix; $\mathrm{Diag}\{\cdot\}$ be a diagonal matrix created by extracting the main diagonal elements of a square matrix or the diagonalization of a vector and $\mathrm{vec}(\cdot)$ for a operator that vectorizes a matrix by stacking its columns vertically.

### 2.1 The restricted multivariate skew normal distribution

A $p$-dimensional random vector $Y$ is said to follow a rMSN distribution with location vector $\mu\in\mathbb{R}^p$, scale covariance matrix $\Sigma$ and skewness vector $\lambda\in\mathbb{R}^p$ if its pdf is

$$f(y) \quad = \quad 2\,\phi_p(y;\mu,\Omega)\Phi\big((1-\lambda^{\mathrm{T}}\Omega^{-1}\lambda)^{-1/2}\lambda^{\mathrm{T}}\Omega^{-1}(y-\mu)\big). \tag{1}$$

where $\Omega = \Sigma + \lambda\lambda^{\mathrm{T}}$. In usual notation, we shall write $Y \sim rSN_p(\mu, \Sigma, \lambda)$ for a random vector with density (1). If $\lambda = 0$, the density of $Y$ will be reduced to $N_p(\mu, \Sigma)$ density.

Based on Pyne et al. (2009), the rMSN distribution can be obtained by a simple convolution-type stochastic representation, given by

$$Y = \mu + \lambda|X_0| + X_1, \quad X_0 \perp X_1, \tag{2}$$

where $X_0 \sim N(0,1)$, $X_1 \sim N_p(0, \Sigma)$ and the symbol "$\perp$" indicates independence. Pyne et al. (2009) formulated a novel mixture model based on this characterization of skew-normal distribution. There exists various alternative extensions of the (univariate) skew normal distribution (Azzalini, 1985), including the multivariate skew normal distributions given by Azzalini and Dalla Valle (1996), Branco and Dey (2001), and Lachos et al. (2010) These models are closely related and are identical after a reparameterization; see Lee and Mclachlan (2013b).

Writing $V = |X_0|$, a hierarchical formulation of (2) can be represented as

$$Y \mid (V = v) \sim N_p(\mu + \lambda v, \Sigma) \text{ and } \gamma \sim TN\big(0, 1\,; (0, \infty)\big). \tag{3}$$

As a consequence, the expectation and covariance matrix of $Y$ are

$$E(Y) = \mu + \lambda\Big(\frac{2}{\pi}\Big)^{1/2} \text{ and } \mathrm{cov}(Y) = \Sigma + \Big(1 - \frac{2}{\pi}\Big)\lambda\lambda^{\mathrm{T}}. \tag{4}$$

The result follows immediately from the first two moments of truncated normal distributions (Lin et al., 2007, Lemma 2) and the law of iterated expectations.

## 2.2 The restricted multivariate skew $t$-distribution

Formally, a $p$-dimensional random vector $Y$ is said to follow a rMST with location vector $\mu \in \mathbb{R}^p$, scale covariance matrix $\Sigma$, skewness vector $\lambda \in \mathbb{R}^p$ and df $\nu \in (0, \infty)$, denoted as $rSt_p(\mu, \Sigma, \lambda, \nu)$, if it can be represented by

$$\begin{aligned} Y &= \mu + W^{-1/2}X, \; X \sim rSN_p(0, \Sigma, \lambda), \\ W &\sim \mathrm{gamma}(\nu/2, \nu/2), \quad X \perp W, \end{aligned} \tag{5}$$

where gamma$(\alpha, \beta)$ stands for a gamma distribution with mean $\alpha/\beta$. If $\lambda = 0$, the distribution of $Y$ reduces to $t_p(\mu, \Sigma, \nu)$ and to $rSN_p(\mu, \Sigma, \lambda)$ as $\nu \to \infty$. In addition, this class of distributions also includes the multivariate normal distribution, recovered by setting $\lambda = 0$ and $\nu \to \infty$. Combining the strengths of the MVT and rMSN distributions, the rMST distribution offers a robustness mechanism against both asymmetry and outliers observed in the data.

**Proposition 1.** *Let $X \mid w \sim N_p(\mu, w^{-1}\Sigma)$ and $W \sim$ gamma$(\alpha, \beta)$, then the joint pdf of $X$ and $w$ has the following relationship:*

$$\begin{aligned} & \phi_p(x; \mu, w^{-1}\Sigma) f_G(w; \alpha, \beta) \\ = \; & t_p(x; \mu, \alpha^{-1}\beta\Sigma, 2\alpha) f_G\Big(w; \frac{2\alpha + p}{2}, \frac{2\beta + \Delta}{2}\Big), \end{aligned} \tag{6}$$

*where $f_G(\cdot; \alpha, \beta)$ denotes the pdf of a gamma distribution with mean $\alpha/\beta$ and the quadratic form $\Delta = (x - \mu)^{\mathrm{T}}\Sigma^{-1}(x - \mu)$ is called the Mahalanobis squared distance.*

**Proof:** See Supplementary Appendix A.

**Proposition 2.** *For any scalar $a \in \mathbb{R}$,*

$$E\big(\Phi(aW^{1/2})\big) = T\Big(a\Big(\frac{\alpha}{\beta}\Big)^{1/2}; 2\alpha\Big).$$

**Proof:** The result is a special case of Lin (2010, Appendix A) when $p = 1$.

From (5), it is clear that the rMST distribution corresponds to a two-level hierarchical representation

$$Y \mid (W = w) \;\sim\; rSN_p\big(\mu, w^{-1}\Sigma, w^{-1/2}\lambda\big) \text{ and } W \sim \text{gamma}(\nu/2, \nu/2). \tag{7}$$

By Propositions 1 and 2, integrating $W$ from the joint density of $(Y, W)$ yields the marginal density of $Y$

$$f(y) \;=\; 2t_p(y; \mu, \Omega, \nu)\; T\Big(A\Big(\frac{\nu + p}{\nu + M}\Big)^{1/2}; \nu + p\Big), \tag{8}$$

where $\Omega = \Sigma + \lambda\lambda^{\mathrm{T}}$, $A = (1-\lambda^{\mathrm{T}}\Omega^{-1}\lambda)^{-1/2}\lambda^{\mathrm{T}}\Omega^{-1}(y-\mu)$ and $M = (y-\mu)^{\mathrm{T}}\Omega^{-1}(y-\mu)$. Supplementary Figure 1 presents various scatter diagrams and contours together with their histograms of the density associated with a bivariate rMST distribution, where $\mu = 0$, $\Sigma = \begin{bmatrix} 1 & \rho \\ \rho & 1 \end{bmatrix}$, $\lambda = (\lambda_1, \lambda_1)^{\mathrm{T}}$ and $\nu = 4$. The values of $\lambda_1$, $\lambda_2$ and $\rho$ reveal different shapes of bivariate rMST distributions. It is apparent that these plots are not elliptical and can be highly correlated and skewed toward different directions depending on their choices of parameters.

Using (4) and the law of iterated expectations, the mean and covariance matrix of $Y$ are

$$E(Y) \quad = \quad \mu + \Big(\frac{\nu}{\pi}\Big)^{1/2} \frac{\Gamma(\frac{\nu-1}{2})}{\Gamma(\frac{\nu}{2})}\lambda \tag{9}$$

and

$$\mathrm{cov}(Y) \quad = \quad \frac{\nu}{\nu-2}\Omega - \frac{\nu}{\pi}\Big\{\frac{\Gamma\big((\nu-1)/2\big)}{\Gamma(\nu/2)}\Big\}^2 \lambda\lambda^{\mathrm{T}}. \tag{10}$$

The mean and covariance matrix do not exist when $\nu \leq 1$ and $\nu \leq 2$, respectively.

**Proposition 3.** *If* $Y \sim rSt_p(\mu, \Sigma, \lambda, \nu)$, *then*

$$\frac{(Y-\mu)^{\mathrm{T}}\Omega^{-1}(Y-\mu)}{p} \quad \sim \quad F_{p,\nu}.$$

**Proof:** See Supplementary Appendix B.

# 3 Skew-$t$ factor analysis model

## 3.1 Model formulation

Suppose that $Y = \{Y_1, \cdots, Y_n\}$ constitutes a random sample of $n$ $p$-dimensional observations. To improve the robustness for modelling correlation in presence of asymmetric levels of sources, we consider a generalization of the $t$FA model in which the latent factor is described by the rMST distribution defined in (8). The model

considered here is

$$
\begin{aligned}
Y_j &= \mu + BU_j + \varepsilon_j \quad \text{with} \\
\begin{bmatrix} U_j \\ \varepsilon_j \end{bmatrix} &\sim rSt_{q+p}\left( \begin{bmatrix} -a_\nu \Lambda^{-1/2}\lambda \\ 0 \end{bmatrix}, \begin{bmatrix} \Lambda^{-1} & 0 \\ 0 & D \end{bmatrix}, \begin{bmatrix} \Lambda^{-1/2}\lambda \\ 0 \end{bmatrix}, \nu \right), \qquad (11)
\end{aligned}
$$

for $j = 1, \dots, n$, where $\mu$ is a $p$-dimensional location vector, $B$ is a $p \times q$ matrix of factor loadings, $U_j$ is a $q$-dimensional vector ($q < p$) of latent variables called *factors*, $\varepsilon_j$ is a $p$-dimensional vector of errors called *specific factors*, $D$ is a positive diagonal matrix, $\Lambda = I_q + (1 - a_\nu^2(\nu-2)/\nu)\,\lambda\lambda^{\mathrm{T}}$ with

$$
a_\nu = (\nu/\pi)^{1/2} \frac{\Gamma\left((\nu-1)/2\right)}{\Gamma\left(\nu/2\right)} \qquad (12)
$$

being a scaling coefficient. An appealing feature of model (11) is that

$$
E(U_j) = 0 \quad \text{and} \quad \mathrm{cov}(U_j) = \{\nu/(\nu-2)\} I_q,
$$

which coincide with the conditions under the $t$FA model. According to (7), the STFA model has a two-level hierarchical representation:

$$
Y_j \mid w_j \sim rSN_p(\mu - a_\nu \alpha, w_j^{-1}\Sigma, w_j^{-1/2}\alpha) \text{ and } W_j \sim \mathrm{gamma}(\nu/2, \nu/2). \qquad (13)
$$

Derivation of the marginal distribution of $Y$ can be accomplished by direct calculation which leads to

$$
Y_j \sim rSt_p(\mu - a_\nu\alpha, \Sigma, \alpha, \nu),
$$

where $\Sigma = B\Lambda^{-1}B^{\mathrm{T}} + D$ and $\alpha = B\Lambda^{-1/2}\lambda$. The marginal density of $Y_j$ is

$$
f(y_j;\theta) = 2t_p(y_j; \mu - a_\nu\alpha, \Omega, \nu) T\Big( A_j \Big( \frac{\nu+p}{\nu+M_j} \Big)^{1/2}; \nu + p \Big), \qquad (14)
$$

where $\Omega = \Sigma + \alpha\alpha^{\mathrm{T}}$, $M_j = (y_j - \mu + a_\nu\alpha)^{\mathrm{T}}\Omega^{-1}(y_j - \mu + a_\nu\alpha)$ and $A_j = h_j/\sigma$ with $h_j = \alpha^{\mathrm{T}}\Omega^{-1}(Y_j - \mu + a_\nu\alpha)$ and $\sigma^2 = 1 - \alpha^{\mathrm{T}}\Omega^{-1}\alpha$.

It therefore follows from (9) and (10) that

$$
E(Y_j) = \mu \quad \text{and} \quad \mathrm{cov}(Y_j) = \frac{\nu}{\nu-2}(BB^{\mathrm{T}} + D).
$$

With the STFA model (11), the parameters estimates of $\mu$, $B$, $D$ and $\nu$ can be used to recover the sample mean and sample covariance for both the $t$FA and STFA models in such a way that the models are likely comparable. Like the traditional FA models, the STFA model enjoys the scale invariance property (Anderson, 2003) and can be reduced to $t$FA model by imposing zero skewness for $U_j$.

For a hidden dimensionality $q > 1$, the STFA model also suffers from an identifiability problem associated with the rotation invariance of loading matrix $B$, since model (11) still satisfies when $B$ is replaced by $BR$, where $R$ is any orthogonal rotation matrix of order $q$. To remedy the situation of rotational indeterminacy, there are several different ways of placing rotational identifiability constraints. The most popular method is to choose $R$ such that $B^{\mathrm{T}} D^{-1} B$ is a diagonal matrix (Lawley and Maxwell, 1971) with its diagonal elements arranged in a descending order. The other commonly used technique is to constrain the loading matrix $B$ so that the upper-right triangle is zero and the diagonal entries are strictly positive (e.g., Fokoué and Titterington, 2003; Lopes and West, 2004). Both methods impose $q(q-1)/2$ constraints on $B$. Therefore, the number of free parameters to be estimated is $m = p(q+2) + q - q(q-1)/2 + 1$.

### 3.2 Maximum likelihood estimation via the ECM algorithm

To help the derivation of the algorithm, we adopt the following scaling transformation:

$$\tilde{B} \triangleq B\Lambda^{-1/2} \quad \text{and} \quad \tilde{U}_j \triangleq \Lambda^{1/2} U_j.$$

Clearly, the model remains invariant under the above transformation. It follows from (3) and (13) that the STFA model can be formulated in a flexible hierarchical representation as follows:

$$
\begin{aligned}
Y_j \mid (\tilde{U}_j, v_j, w_j) &\sim N_p(\mu + \tilde{B}\tilde{U}_j, w_j^{-1}D),\\
\tilde{U}_j \mid (v_j, w_j) &\sim N_q\big((v_j - a_\nu)\lambda, w_j^{-1}I_q\big),\\
V_j \mid w_j &\sim TN\big(0, w_j^{-1}; (0,\infty)\big),\\
W_j &\sim \text{gamma}(\nu/2, \nu/2). \qquad (15)
\end{aligned}
$$

Consequently, applying Bayes' rule, it suffices to show

$$
\begin{aligned}
\tilde{U}_j \mid (y_j, v_j, w_j) &\sim N_q\big(q_j, w_j^{-1}C\big),\\
V_j \mid (y_j, w_j) &\sim TN\big(h_j, w_j^{-1}\sigma^2; (0,\infty)\big),\\
f(w_j; y_j) &= \frac{\Phi(w_j^{1/2}A_j)}{T\Big(A_j(\frac{\nu+p}{\nu+M_j})^{1/2}; \nu+p\Big)} f_G\Big(w_j; \frac{\nu+p}{2}, \frac{\nu+M_j}{2}\Big), \qquad (16)
\end{aligned}
$$

where $q_j = C\big\{d_j + \lambda(v_j - a_\nu)\big\}$, $\quad d_j = \tilde{B}^{\mathrm{T}}D^{-1}(Y_j - \mu)$ and $C = (I_q + \tilde{B}^{\mathrm{T}}D^{-1}\tilde{B})^{-1}$.

In what follows, define $c_j(r) = \{(\nu + p + r)/(M_j + \nu)\}^{1/2}$, where $r = -2, 0, 2$. We establish the following proposition, which is crucial for the calculation of some conditional expectations involved in the proposed ECM algorithm.

**Proposition 4.** *From (16), we have the following conditional expectations:*

$$
\begin{aligned}
E(W_j \mid y_j) &= \{c_j(0)\}^2 \frac{T(A_jc_j(2); \nu+p+2)}{T(A_jc_j(0); \nu+p)}, \qquad (17)\\
E(V_j \mid y_j) &= h_j + \sigma c_j(-2)\frac{t(A_jc_j(-2); \nu+p-2)}{T(A_jc_j(0); \nu+p)}, \qquad (18)\\
E(W_jV_j \mid y_j) &= h_jE(W_j \mid y_j) + \sigma c_j(0)\frac{t(A_jc_j(0); \nu+p)}{T(A_jc_j(0); \nu+p)}, \qquad (19)\\
E(W_jV_j^2 \mid y_j) &= \sigma^2 + h_jE(W_jV_j \mid y_j), \qquad (20)\\
E(W_j\tilde{U}_j \mid y_j) &= C\Big[d_jE(W_j \mid y_j) + \lambda\Big\{E(W_jV_j \mid y_j) - a_\nu E(W_j \mid y_j)\Big\}\Big], \qquad (21)\\
E(W_jV_j\tilde{U}_j \mid y_j) &= C\Big[d_jE(W_jV_j \mid y_j)\\
&\quad + \lambda\Big\{E(W_jV_j^2 \mid y_j) - a_\nu E(W_jV_j \mid y_j)\Big\}\Big], \qquad (22)
\end{aligned}
$$

*and*

$$E(W_j\tilde{U}_j\tilde{U}_j^{\mathrm{T}} \mid y_j) = \Big[\Big\{E(W_jV_j\tilde{U}_j \mid y_j) - a_\nu E(W_j\tilde{U}_j \mid y_j)\Big\}\lambda^{\mathrm{T}} + E(W_j\tilde{U}_j \mid y_j)d_j^{\mathrm{T}} + I_q\Big]C. \quad (23)$$

**Proof:** See Supplementary Appendix C.

The expectation-maximization (EM) algorithm (Dempster et al., 1977) is a popular iterative method to compute the ML estimates when the data are incomplete or of latent variables. Given an initial solution $\theta^{(0)}$, the implementation of the EM algorithm consists of alternating repeatedly the Expectation (E)- and Maximization (M)-steps until convergence, e.g., a successive increase of the log-likelihood diminishes. Often in many practical problems, the solution to the M-step may encounter some difficulties such that no closed-form expressions exist for updating parameters. For ML estimation of the STFA model, we resort to the ECM algorithm (Meng and Rubin, 1993) in which the M-step is replaced by a sequence of computationally simper conditional maximization (CM) steps while sharing all appealing advantages of the standard EM algorithm.

For notational convenience, let $y = (y_1^{\mathrm{T}}, \dots, y_n^{\mathrm{T}})^{\mathrm{T}}$ be the observed data. Moreover, we let $U = (U_1^{\mathrm{T}}, \dots, U_n^{\mathrm{T}})^{\mathrm{T}}$, $V = (V_1, \dots, V_n)^{\mathrm{T}}$ and $W = (W_1, \dots, W_n)^{\mathrm{T}}$, which are treated as missing values in the complete data framework. In light of (15), the complete data log-likelihood function for $\theta = (\mu, B, D, \lambda, \nu)$ given $y_c = (y^{\mathrm{T}}, U^{\mathrm{T}}, V^{\mathrm{T}}, W^{\mathrm{T}})^{\mathrm{T}}$, aside from additive constants, is

$$\begin{aligned}\ell_c(\theta; y_c) &= -\frac{n}{2}\log \mid D \mid -\frac{1}{2}\mathrm{tr}\Big(D^{-1}\sum_{j=1}^{n}\tilde{\Upsilon}_j\Big) \\ &\quad -\frac{1}{2}\sum_{j=1}^{n}\Big[W_j\Big\{(V_j - a_\nu)^2\lambda^{\mathrm{T}}\lambda - 2(V_j - a_\nu)\lambda^{\mathrm{T}}\tilde{U}_j + \tilde{U}_j\tilde{U}_j^{\mathrm{T}}\Big\}\Big] \\ &\quad +\frac{n\nu}{2}\log\Big(\frac{\nu}{2}\Big) - n\log\Gamma\Big(\frac{\nu}{2}\Big) + \frac{\nu}{2}\sum_{j=1}^{n}(\log W_j - W_j), \quad (24)\end{aligned}$$

where $\tilde{\Upsilon}_j = W_j(y_j - \mu - \tilde{B}\tilde{U}_j)(y_j - \mu - \tilde{B}\tilde{U}_j)^{\mathrm{T}}$.

To calculate the expected complete data log-likelihood, called the $Q$-function, it involves the calculation of the following conditional expectations:

$$\begin{aligned}
\hat{w}_j^{(k)} &= E(W_j \mid y_j, \hat{\theta}^{(k)}), \quad \hat{\kappa}_j^{(k)} = E(\log W_j \mid y_j, \hat{\theta}^{(k)}), \\
\hat{s}_{1j}^{(k)} &= E(W_j V_j \mid y_j, \hat{\theta}^{(k)}), \quad \hat{s}_{2j}^{(k)} = E(W_j V_j^2 \mid y_j, \hat{\theta}^{(k)}), \quad \hat{\tilde{\Omega}}_j^{(k)} = E(W_j \tilde{U}_j \tilde{U}_j^{\mathrm{T}} \mid y_j, \hat{\theta}^{(k)}), \\
\hat{\tilde{\eta}}_j^{(k)} &= E(W_j \tilde{U}_j \mid y_j, \hat{\theta}^{(k)}) \text{ and } \hat{\tilde{\zeta}}_j^{(k)} = E(W_j V_j \tilde{U}_j \mid y_j, \hat{\theta}^{(k)}), \qquad (25)
\end{aligned}$$

which are directly obtainable from using (17)-(23) given in Proposition 4. As a result, the $Q$-function can be written as

$$\begin{aligned}
Q(\theta; \hat{\theta}^{(k)}) &= -\frac{n}{2} \log \mid D \mid - \frac{1}{2} \mathrm{tr}\Big(D^{-1} \sum_{j=1}^{n} \hat{\tilde{\Upsilon}}_j^{(k)}\Big) \\
&\quad - \frac{1}{2} \sum_{j=1}^{n} \Big\{ (\hat{s}_{2j}^{(k)} - 2a_\nu \hat{s}_{1j}^{(k)} + a_\nu^2 \hat{w}_j^{(k)}) \lambda^{\mathrm{T}} \lambda - 2\lambda^{\mathrm{T}} (\hat{\tilde{\zeta}}_j^{(k)} - a_\nu \hat{\tilde{\eta}}_j^{(k)}) + \hat{\tilde{\Omega}}_j^{(k)} \Big\} \\
&\quad + \frac{n\nu}{2} \log\Big(\frac{\nu}{2}\Big) - n \log \Gamma\Big(\frac{\nu}{2}\Big) + \frac{\nu}{2} \sum_{j=1}^{n} (\hat{\kappa}_j^{(k)} - \hat{w}_j^{(k)}), \qquad (26)
\end{aligned}$$

where

$$\begin{aligned}
\hat{\tilde{\Upsilon}}_j^{(k)} &= \hat{w}_j^{(k)} (y_j - \mu)(y_j - \mu)^{\mathrm{T}} - \tilde{B} \hat{\tilde{\eta}}_j^{(k)} (y_j - \mu)^{\mathrm{T}} - (y_j - \mu) \hat{\tilde{\eta}}_j^{(k)\mathrm{T}} \tilde{B}^{\mathrm{T}} \\
&\quad + \tilde{B} \hat{\tilde{\Omega}}_j^{(k)} \tilde{B}^{\mathrm{T}}, \qquad (27)
\end{aligned}$$

which contains free parameters $\mu$ and $\tilde{B}$. In summary, the implementation of the ECM algorithm proceeds as follows:

**E-step:** Given $\theta = \hat{\theta}^{(k)}$, compute $\hat{w}_j^{(k)}$, $\hat{\kappa}_j^{(k)}$, $\hat{s}_{1j}^{(k)}$, $\hat{s}_{2j}^{(k)}$, $\hat{\tilde{\eta}}_j^{(k)}$, $\hat{\tilde{\zeta}}_j^{(k)}$ and $\hat{\tilde{\Omega}}_j^{(k)}$ in (25), for $j = 1, \ldots, n$.

**CM-step 1:** Update $\hat{\mu}^{(k)}$ by maximizing (26) over $\mu$, which leads to

$$\hat{\mu}^{(k+1)} = \frac{\sum_{j=1}^{n} \Big( \hat{w}_j^{(k)} y_j - \hat{\tilde{B}}^{(k)} \hat{\tilde{\eta}}_j^{(k)} \Big)}{\sum_{j=1}^{n} \hat{w}_j^{(k)}}.$$

**CM-step 2:** Given $\mu = \hat{\mu}^{(k+1)}$, update $\hat{\tilde{B}}^{(k)}$ by maximizing (26) over $\tilde{B}$, which gives

$$\hat{\tilde{B}}^{(k+1)} = \Big\{ \sum_{j=1}^{n} \big(y_j - \hat{\mu}^{(k+1)}\big)\hat{\tilde{\eta}}_j^{(k)\mathrm{T}} \Big\} \Big( \sum_{j=1}^{n} \hat{\tilde{\Omega}}_j^{(k)} \Big)^{-1}.$$

**CM-step 3:** Given $\mu = \hat{\mu}^{(k+1)}$ and $\tilde{B} = \hat{\tilde{B}}^{(k+1)}$, update $\hat{D}^{(k)}$ by maximizing (26) over $D$, which leads to

$$\hat{D}^{(k+1)} = \frac{1}{n}\mathrm{Diag}\Big( \sum_{j=1}^{n} \hat{\tilde{\Upsilon}}_j^{(k)} \Big).$$

where $\hat{\tilde{\Upsilon}}_j^{(k)}$ is $\tilde{\Upsilon}_j^{(k)}$ in (27) with $\mu$ and $\tilde{B}$ replaced by $\hat{\mu}^{(k+1)}$ and $\hat{\tilde{B}}^{(k+1)}$, respectively.

**CM-step 4:** Update $\hat{\lambda}^{(k)}$ by maximizing (26) over $\lambda$, which gives

$$\hat{\lambda}^{(k+1)} = \frac{\hat{\tilde{\zeta}}_j^{(k)} - a_\nu \hat{\tilde{\eta}}_j^{(k)}}{\hat{s}_{2j}^{(k)} - 2a_\nu \hat{s}_{1j}^{(k)} + a_\nu^2 \hat{w}_j^{(k)}}.$$

**CM-step 5:** Calculate $\hat{\nu}^{(k+1)}$ by maximizing (26) over $\nu$, which is equivalent to solve the root of the following equation:

$$\begin{aligned} -\frac{1}{n}\sum_{j=1}^{n} \Big\{ (-2a_\nu' \hat{s}_{1j}^{(k)} + 2a_\nu' a_\nu \hat{w}_j^{(k)})\lambda^{\mathrm{T}}\lambda + 2a_\nu' \lambda^{\mathrm{T}} \hat{\tilde{\eta}}_j^{(k)} \Big\} & \\ + \log\left(\frac{\nu}{2}\right) - \mathrm{DG}\Big(\frac{\nu}{2}\Big) + 1 + \frac{1}{n}\sum_{j=1}^{n} (\hat{\kappa}_j^{(k)} - \hat{w}_j^{(k)}) &= 0. \end{aligned}$$

where DG denotes the digamma function and

$$a_\nu' = \frac{da_\nu}{d\nu} = \frac{1}{2}\left(\frac{1}{\pi\nu}\right)^{1/2} \frac{\Gamma\Big(\frac{\nu-1}{2}\Big)}{\Gamma\Big(\frac{\nu}{2}\Big)} + 2\left(\frac{\nu}{\pi}\right)^{1/2} \frac{\Gamma\Big(\frac{\nu-1}{2}\Big)}{\Gamma\Big(\frac{\nu}{2}\Big)} \Big\{ \mathrm{DG}\Big(\frac{\nu-1}{2}\Big) - \mathrm{DG}\Big(\frac{\nu}{2}\Big) \Big\}.$$

In the above CM-step 5, the R function 'uniroot' is emplyed to obtain the solution of $\nu$. To facilitate faster convergence, the range of $\nu$ is restricted to have a maximum of 200, which does not affect the inference when the underlying distribution of factor scores are near-normality. Upon convergence, the ML estimate of $\theta$ is denoted by $\hat{\theta} = (\hat{\mu}, \hat{B}, \hat{D}, \hat{\lambda})$, where $\hat{B} = \hat{\tilde{B}}\hat{\Lambda}^{1/2}$ and $\hat{\Lambda} = I_q + (1 - \frac{\hat{\nu}-2}{\hat{\nu}}\hat{a}_\nu^2)\hat{\lambda}\hat{\lambda}^{\mathrm{T}}$. Consequently, the estimation of factor scores through *conditional prediction* is obtained by

$$\hat{U}_j = E(U_j \mid y_j, \hat{\theta}) = \hat{\Lambda}^{-1/2}\hat{C}\Big\{\hat{d}_j + \hat{\lambda}\big(\hat{v}_j - \hat{a}_\nu\big)\Big\},$$

where $\hat{v}_j = E(V_j \mid y_j, \hat{\theta})$ can be evaluated via (18) with $\theta$ replaced by $\hat{\theta}$ and $\hat{a}_\nu$ is $a_\nu$ in (12) with $\nu$ replaced by $\hat{\nu}$.

We further make some remarks on the implementation of the proposed ECM algorithm.

**Remark 1.** To monitor the convergence based on the monotonicity property of the algorithm, a simple way is to repeat iterations after a certain number of iterations, say $K$, or until the difference between two successive log-likelihood evaluations is small enough, say $\ell^{(k+1)} - \ell^{(k)} < \epsilon$, where for brevity of notation $\ell^{(k)}$ means the log-likelihood value evaluated at $\hat{\theta}^{(k)}$ and $\epsilon$ is a user-specified tolerance. In our analysis, we use $K = 5,000$ and $\epsilon = 10^{-6}$.

**Remark 2.** As analogous to other iterative optimization procedures, one needs to search for appropriate initial values to avoid divergence or time-consuming computation. A direct way of deriving the initial estimate for mean vector, factor loading and error covariance matrix can be obtained by performing a simple FA fit using the *factanal* command in the R package. The resulting estimates are taken as initial values, namely $\hat{\mu}^{(0)}$, $\hat{B}^{(0)}$ and $\hat{D}^{(0)}$, respectively. Next, compute the factor scores via the conditional prediction method. The initial skewness vector $\hat{\lambda}^{(0)}$ and df $\hat{\nu}^{(0)}$ are obtained by fitting the rMST distribution to the sample of factor scores via the R package EmSkew (Wang et al., 2009).

**Remark 3.** For model selection and determination of $q$, the fitting results are compared based on the Akaike's information criterion (AIC; Akaike, 1973) and the Bayesian information criterion (BIC; Schwarz, 1978), which are defined as

$$\text{AIC} = 2m - 2\ell_{\max} \qquad \text{and} \qquad \text{BIC} = m\log n - 2\ell_{max}.$$

where $\ell_{max}$ is the maximized log-likelihood and $m$ is the number of free parameters in the considered model.

## 4 Provision of standard errors

Under regularity conditions (Zacks, 1971), the asymptotic covariance matrix of $\hat{\theta}$ can be approximated by the inverse of the observed information matrix; see also Efron and Hinkley (1978). Specifically, the observed information matrix is defined as the Hessian of the negative of the log-likelihood function

$$I(\hat{\theta}; y) = -\frac{\partial^2 \ell(\theta; y)}{\partial\theta\partial\theta^{\mathrm{T}}}\Big|_{\theta=\hat{\theta}}.$$

To obtain $I(\hat{\theta}; y)$ numerically, Jamshidian (1997) suggested using the central difference method. Let $G = [g_1; \cdots \mid g_m]$ be a $m \times m$ matrix with the $c$th column being

$$g_c = \frac{s(\theta + h_c e_c; y) - s(\theta - h_c e_c; y)}{2h_c}, \quad c = 1, \cdots, m,$$

where $s(\theta; y) = \partial\ell(\theta; y)/\partial\theta$ is the score vector of $\ell(\theta; y)$, $e_c$ is a unit vector with all of its elements equal to zero except for its $c$th element which is equal to 1, $h_c$ is a small number, and $m$ is the number of parameters in $\theta$. Explicit expressions for the elements of $s(\theta; y)$ are summarized in Supplementary Appendix D.

Since $G$ may not be symmetric, we suggest using

$$\tilde{I}(\hat{\theta}; y) \;=\; -\frac{G + G^{\mathrm{T}}}{2}. \tag{28}$$

to approximate $I(\hat{\theta}; y)$ The asymptotic standard errors of $\hat{\theta}$ can be calculated by taking the square roots of the diagonal elements of $[\tilde{I}(\hat{\theta}; y)]^{-1}$.

Notably, the inverse of (28) is not always guaranteed to yield proper (positive) standard errors. The parametric bootstrap method (Efron and Tibshirani, 1993), although computationally expensive, is often used instead to obtain estimates of the standard errors. Let $\hat{f}(y; \hat{\theta})$ be the estimated density function of (14) obtained from fitting the STFA model to the original data. Obtaining bootstrap standard error estimates consists of the following four steps.

1. Drawing a bootstrap sample $y_1^*, \dots, y_n^*$ from the fitted distribution $\hat{f}(y; \hat{\theta})$.

2. Compute the ML estimates $\hat{\theta}^*$ from fitting the STFA model to the generated bootstrap samples $y_1^*, \dots, y_n^*$.

3. Repeat Steps 1 and 2 a large number of times, say $B$, thereby obtaining bootstrap replications, namely $\hat{\theta}_1^*, \dots, \hat{\theta}_B^*$.

4. Estimate the bootstrap standard errors of $\hat{\theta}$ via the sample standard errors of $\hat{\theta}_1^*, \dots, \hat{\theta}_B^*$.

## 5 A numerical illustration

As an illustration, we apply the proposed technique to the Australian Institute of Sport (AIS) data, which were originally reported by Cook and Weisberg (1994) and subsequently analyzed by Azzalini and Dalla Valle (1996), Azzalini and Capitaino (1999, 2003) and Azzalini (2005), among others. The dataset consists of $p = 11$ physical and hematological measurements on athletes in different sports which are almost equally bisected between 102 male and 100 female.

Table 1 about here

For simplicity of illustration, we focus solely on $n = 102$ observations of male. A summary of 11 attributes along with their sample skewness and kurtosis is given in Table 1. It is readily seen that most of attributes are noticeable moderately to strongly skewed and leptokurtotic in nature.

Figure 1 about here

Figure 1 depicts the histograms and corresponding normal quantile plots of the first three factor score estimates obtained from the classical FA with $q = 4$. The histograms in the left panels indicate that the distribution of factor scores deviate from normality due to positive skewness and high excess kurtosis. The feature can also be demonstrated through the normal quantile-quantile plots shown in the right panels. This motivates us to advocate the use of STFA model as a proper tool for the analysis of this data set.

Table 2 about here

Next, we are interested in comparing the ML results of STFA with those obtained under three reduced models, namely the FA, $t$FA and SNFA. The data have been standardized to have zero mean and unit standard deviation to avoid variables having a greater impact due to different scales. We fit these models with $q$ ranging from 1 to 6 using the ECM algorithm developed in Section 3. Notice that the choice of maximum $q = 6$ satisfies the restriction $(p-q)^2 \leq (p+q)$ as suggested by Eq. (8.5) of McLachlan and Peel (2000).

Table 3 about here

A summary of ML fitting results, including the maximized log-likelihood values, the number of parameters together with the AIC and BIC values, is reported in Table 2. Judging from the table, the best fitted model is STFA with $q = 4$, no matter which selection criterion was used. Table 3 reports the ML solutions of the best chosen model along with the standard errors in parentheses obtained using 500 bootstrap replications. We found that the estimated skewness parameters are moderately to highly significant, revealing that the joint distribution of latent factors are skewed. Moreover, the estimated df ($\hat{\nu} = 6.28$) is quite small, confirming the presence of thick tails.

Observing the unrotated solution of factor loading displayed in the 3-6th columns of Table 3, the first factor can be labelled *general nutritional status*, with a very high loading on lbm, followed by Wt, Ht and bmi. The second factor, which loads heavily on rcc, Hc and Hg, might be called a *hematological factor*. The third factor can be viewed as *overweight assessment indices* since the bmi, ssf and Bfat load highly on this factor. The fourth factor is not easily interpreted at this point.

The comparison process is also conducted for the original (non-standardized) data. Clearly, as shown in Supplementary Figure 2, the STFA still provides the best overall fit, followed by $t$FA and SNFA. The fit of FA is the worst, indicating a lack of adequacy of normality assumptions for this dataset. It is also noted that both criteria prefer four-factor solutions under all scenarios.

Figure 2 about here

We consider diagnostics to assess the validity of the underlying distributional assumption of $Y$. For FA, we can use the Mahalanobis-like distance $(Y-\hat{\mu})^{\mathrm{T}}\hat{\Omega}^{-1}(Y-\hat{\mu})$, which has an asymptotic chi-square distribution with $p$ df. Checking the normality assumption can be achieved by constructing the Healy's (1968) plot. To further assess the fitness of STFA, it follws from Proposition 11 that $(Y-\hat{\mu})^{\mathrm{T}}\hat{\Omega}^{-1}(Y-\hat{\mu})/p$ follows the $F$ distribution with $p$ and $\nu$ dfs. In this case, one can construct another Healy-type plot (or the Snedecor-F plot) by plotting the nominal values $(1/n, 2/n, \ldots, 1)$ against the empirical cdf values of the ordered $F$ statistics. As such, one can examine whether the corresponding Healy's plot resembles a straight line through the origin having unit slope. In other words, the greater the departure from the 45-degree line, the greater the evidence for concluding a poor fitting of the model. Inspecting Healy's plots shown in Figure 2, the STFA adapts the identity more closely than does the FA, suggesting that it is appropriate to use a heavy-tailed distribution.

Figure 3 about here

Figure 3 depicts coordinate projected scatter plots for each pair of four selected variables superimposed with the marginal contours obtained by marginalization of the best fitted STFA model. A visual inspection reveals that the fitted contours adapt the shape of the scattering pattern satisfactorily. To summarize, the implementation of STFA procedure tends to be more reasonable for analyzing this data set.

## 6 Conclusion

We introduce an extension of FA models obtained by replacing the normality assumption for the latent factors and errors with a joint rMST distribution, called the STFA model, as a new robust tool for dimensionality reduction. The model accommodates both asymmetry and heavy tails jointly and allows practitioners for analyzing data in a wide variety of considerations. We have described a four-level hierarchical representation for the STFA model and presented an analytically simple ECM algorithm for ML estimation in a flexible complete-data framework. We demonstrate our approach with a real data set and show that the STFA model may provide better performance than several existing competitors.

In the situation with the occurrence of missing data, our algorithm can be easily modified to account for missingness based on the scheme proposed in Lin et al. (2006). Due to recent advances in computer power and availability, it is worthwhile to develop Markov chain Monte Carlo (MCMC) methods (Lin et al., 2009 and Lin and Lin, 2011) for carrying out Bayesian inference of the STFA model. It is also of interest to consider a finite mixture representation of STFA models. Our initial work on the latter problem has been limited to mixtures of factors with a skew-normal distribution (Lin et al., 2013).

Also, it should be noted in other unpublished work involving mixtures of factor models (Murray et al., 2013a; 2013b) that the skew $t$-distribution adopted is different to the skew $t$-distribution considered in our paper. Rather it is the limiting form of the generalized hyperbolic distribution, which has some quite different properties. For example, it has one exponential tail and one polynomial tail instead of two polynomial tails as with the usual skew $t$-distribution. Also, as the skewness parameters in its formulation tend to zero, it does not become a skew normal distribution; that is, it does not nest the skew normal distribution as a special case. The unrestricted skew $t$-distribution is considered in Murray et al. (2013c). But as in Murray et al. (2013a), the factor analytic representation applies only to the error terms in the presence of the skewing variables and not the factors.

## References

Akaike, H. (1973) Information theory and an extension of the maximum likelihood principle. In 2nd Int. Symp. on Information Theory (Edited by B. N. Petrov and F. Csaki), 267–281. Akademiai Kiado, Budapest.

Anderson, T.W. (2003) An Introduction to Multivariate Statistical Analysis, third ed. Wiely, New York.

Azzalini, A. (1985) A class of distributions which includes the normal ones. Scandinacian Journal of Statistics. 12, 171–178.

Azzalini, A. (2005) The skew-normal distribution and related multivariate families. Scand. J. Statist. 32, 159–188.

Azzalini, A. and Capitanio, A. (1999) Statistical applications of the multivariate skew normal distribution. J. Roy. Statist. Soc. Ser. B 61, 579–602.

Azzalini, A. and Capitaino, A. (2003) Distributions generated by perturbation of symmetry with emphasis on a multivariate skew $t$-distribution. J. R. Stat. Soc. Ser. B 65, 367–389.

Azzalini, A. and Dalla Valle, A. (1996) The multivariate skew-normal distribution. Biometrika 83, 715–726.

Azzalini, A. and Genton, M.G. (2008) Robust likelihood methods based on the skew-$t$ and related distributions. Int. Statist. Rev., 76, 106–129.

Baek, J. and McLachlan, G.J. (2011) Mixtures of common $t$-factor analyzers for clustering high-dimensional microarray data. Bioinformatics 27, 1269–1276.

Baek, J., McLachlan, G.J., and Flack, L. (2010) Mixtures of factor analyzers with common factor loadings: applications to the clustering and visualisation of high-dimensional data. IEEE Transactions on Pattern Analysis and Machine Intelligence 32, 1298–1309.

Basilevsky, A. (2008) Statistical Factor Analysis and Related Methods: Theory and Applications. John Wiley & Sons, Inc.

Branco, M.D. and Dey, D.K. (2001) A general class of multivariate skew-elliptical distributions. Journal of Multivariate Analysis., 79, 99–113.

Cook, R.D. and Weisberg, S. (1994) An Introduction to Regression Graphics. Wiley, New York.

Dempster, A.P., Laird, N.M. and Rubin, D.B. (1977) Maximum likelihood from incomplete data via the EM algorithm (with discussion). J. R. Stat. Soc. Ser. B 39, 1–38.

Efron, B. and Hinkley, D.V. (1978) Assessing the accuracy of the maximum likelihood estimator: Observed versus expected Fisher Information (with discussion). Biometrika 65 457–487.

Efron, B. and Tibshirani, R. (1986) Bootstrap method for standard errors, confidence intervals, and other measures of statistical accuracy. Stat. Sci. 1, 54–77.

Fokoué, E. and Titterington, D.M., (2003) Mixtures of factor analyzers. Bayesian estimation and inference by stochastic simulation. Mach. Learning 50, 73–94.

Ghahramani, Z. and Hinton, G. (1997) The EM algorithm for mixtures of factor analyzers. Technical Report CRG-TR-96-1, University of Toronto.

Kotz, S. and Nadarajah, S. (2004) Multivariate $t$ distributions and their applications. Cambridge University Press.

Healy, M.J.R. (1968) Multivariate normal plotting. App. Statist. 17, 157–161.

Ho, H.J., Lin, T.I., Chang, H.H., Haase, H.B., Huang, S. and Pyne, S. (2012) Parametric modeling of cellular state transitions as measured with flow cytometry different tissues. BMC Bioinformatics 13 (Suppl 5):S5

Jamshidian, M. (1997) An EM algorithm for ML factor analysis with missing data. In Berkane, M. (Ed.) Latent Variable Modeling and Applications to Causality, (pp. 247-258). Springer Verlag, New York.

Jones, M.C. and Faddy, M.J. (2003) A skew extension of the $t$-distribution, with applications. J. Roy. Statist. Soc. Ser. B, 65, 159–174.

Lachos, V.H., Ghosh, P., and Arellano-Valle, R.B. (2010) Likelihood based inference for skew normal independent linear mixed models. Statistica Sinica. 20, 303–322.

Lange, K.L., Little, R.J.A. and Taylor, J.M.G. (1989) Robust statistical modeling using the $t$ distribution. J. Amer. Statist. Assoc. 84, 881–896.

Lawley, D.N. and Maxwell, A.E. (1971) Factor analysis as a Statistical Method. 2nd ed., Butterworth, London.

Lin, T.I., McLachlan, G.J., and Lee, S.X. (2013). Extending mixtures of factor models using the restricted multivariate skew-normal distribution. Preprint arXiv:1307.1748.

Lee, S. and McLachlan, G.J. (2013a) Finite mixtures of multivariate skew $t$-distributions: some recent and new results Statist. Comput. DOI 10.1007/s11222-012-9362-4.

Lee, S.X. and McLachlan, G.J. (2013b). On mixtures of skew normal and skew $t$-distributions. Adv. Data Anal. Classif. Doi 10.1007/s11634-013-0132-8.

Lin, T.I., Ho, H.J. and Chen, C.L. (2009) Analysis of multivariate skew normal models with incomplete data. J. Multivariate Anal. 100, 2337–2351.

Lin, T.I., Lee, J.C. and Ho, H.J. (2006) On fast supervised learning for normal mixture models with missing information. Pattern Recognit. 39, 1177–1187.

Lin, T.I. (2010) Robust mixture modeling using multivariate skew $t$ distributions. Stat. Comput. 20, 343–356.

Lin, T.I. and Lin, T.C. (2011) Robust statistical modelling using the multivariate skew $t$ distribution with complete and incomplete data. Stat. Model. 11, 253–277.

Lin, T.I., Lee, J.C. and Ho, H.J. (2006) On fast supervised learning for normal mixture models with missing information, Pattern Recog. 39, 1177–1187.

Lin, T.I., Lee, J.C. and Hsieh, W.J. (2007) Robust mixture modeling using the skew t distribution. Stat. Comput. 17, 81–92.

Lopes, H. F. and West M. (2004) Bayesian model assessment in factor analysis. Stat. Sin. 14, 41–67.

Louis, T.A. (1982) Finding the observed information when using the EM algorithm. J. R. Stat. Soc. Ser. B 44, 226–232.

McLachlan, G.J., Bean, R.W. and Jones, L.B.T. (2007) Extension of the mixture of factor analyzers model to incorporate the multivariate $t$-distribution. Comput. Stat. Data Anal. 51, 5327–5338.

McLachlan, G.J. and Peel, D. (2000) Mixture of factor analyzers. In Proceedings of the Seventeenth International Conference on Machine Learning, P. Langley (Ed.). San Francisco: Morgan Kaufmann, 599–606.

McLachlan, G.J.,Peel, D. and Bean, R.W. (2003) Modelling high-dimensional data by mixtures of factor analyzers. Comput. Stat. Data Anal. 41, 379–388.

Meng, X.L. and Rubin, D.B. (1993) Maximum likelihood estimation via the ECM algorithm: a general framework. Biometrika 80, 267–278.

Murray, P.M., Browne, R.P., and McNicholas, P.D. (2013a). Mixtures of skew-$t$ factor analyzers. Preprint arXiv:1305.4301v2.

Murray, P.M., Browne, R.P., and McNicholas, P.D. (2013b). Mixtures of common skew-$t$ factor analyzers. Preprint arXiv:1307.5558v3.

Murray, P.M., Browne, R.P., and McNicholas, P.D. (2013c). Mixtures of 'unrestricted' skew-$t$ factor analyzers. Preprint arXiv:1310.6224v1.

Pyne S., Hu, X., Wang, K., Rossin, E., Lin, T.I., Maier, L.M., Baecher-Allan, C., McLachlan, G.J., Tamayo, P., Hafler, D.A., De Jager, P.L. and Mesirov, J.P., 2009. Automated high-dimensional flow cytometric data analysis. Proc. Natl. Acad. Sci. USA, 106, 8519–8524.

Rossin, E., Lin, T.I., Ho, H.J., Mentzer, S.J. and Pyne, S. (2011) A framework for analytical characterization of monoclonal antibodies based on reactivity profiles in different tissues. Bioinformatics 27, 2746–2753.

Sahu, S.K., Dey, D.K. and Branco, M.D. (2003) A new class of multivariate skew distributions with application to Bayesian regression models. Can. J. Statist. 31, 129–150.

Schwarz, G. (1978) Estimating the dimension of a model. Ann. Statist. 6, 461–464.

Spearman, C. (1904) General intelligence, objectively determined and measured. Am. J. Psy. 15, 201–292.

Titterington, D.M., Smith, A.F.M. and Markov, U.E. (1985) Statistical analysis of finite mixture distributions. Wiley, New York

Wang, K., McLachlan, G.J., Ng, S.K. and Peel, D. (2009) EMMIX-skew (R package version 1.0-12): EM Algorithm for Mixture of Multivariate Skew Normal/$t$ Distributions.

Wang, W.L. and Lin, T.I. (2013) An efficient ECM algorithm for maximum likelihood estimation in mixtures of t-factor analyzers. Comput. Statist. 28, 751–769.

Zacks, S. (1971) The Theory of Statistical Inference. John Wiley, New York

Zhang, J., Li, J. and Liu, C. (2013) Robust factor analysis using the multivariate $t$-distribution. unpublished manuscript.

Table 1: An overview of 11 attributes of 102 male athletes of the AIS data.

| Attribute | Variable | Description | skewness | kurtosis |
|---|---|---|---|---|
| $x_1$ | rcc | Red cell count | 0.924 | 7.730 |
| $x_2$ | wcc | White cell count | 0.859 | 4.579 |
| $x_3$ | Hc | Hematocrit | 1.489 | 10.374 |
| $x_4$ | Hg | Hemoglobin | 0.974 | 5.312 |
| $x_5$ | Fe | Plasma ferritin concentration | 0.877 | 3.133 |
| $x_6$ | bmi | Body mass index | 1.411 | 5.986 |
| $x_7$ | ssf | Sum of skin folds | 1.386 | 4.789 |
| $x_8$ | Bfat | Body fat percentage | 1.528 | 5.080 |
| $x_9$ | lbm | Lean body mass | 0.274 | 3.621 |
| $x_{10}$ | Ht | Height (cm) | 0.072 | 3.001 |
| $x_{11}$ | Wt | Weight (Kg) | 0.390 | 3.410 |

Table 2: Comparison of ML estimation results on $n = 102$ male athletes.

| Model | $q$ | $\ell_{\max}$ | $m$ | AIC | BIC |
|---|---|---|---|---|---|
| FA | 1 | –1299.84 | 33 | 2665.69 | 2752.31 |
| | 2 | –1139.70 | 43 | 2365.41 | 2478.28 |
| | 3 | –788.86 | 52 | 1681.72 | 1818.22 |
| | 4 | –639.65 | 60 | 1399.30 | 1556.79 |
| | 5 | –637.20 | 67 | 1408.40 | 1584.27 |
| | 6 | –633.82 | 73 | 1413.63 | 1605.25 |
| $t$FA | 1 | –1190.30 | 34 | 2448.60 | 2537.85 |
| | 2 | –1065.03 | 44 | 2218.06 | 2333.56 |
| | 3 | –710.57 | 53 | 1527.14 | 1666.27 |
| | 4 | –590.97 | 61 | 1303.94 | 1464.07 |
| | 5 | –588.78 | 68 | 1313.57 | 1492.06 |
| | 6 | –586.09 | 74 | 1320.19 | 1514.44 |
| SNFA | 1 | –1299.76 | 34 | 2667.52 | 2756.77 |
| | 2 | –1135.21 | 45 | 2360.42 | 2478.54 |
| | 3 | –761.47 | 55 | 1632.95 | 1777.32 |
| | 4 | –609.62 | 64 | 1347.24 | 1515.24 |
| | 5 | –606.93 | 72 | 1357.87 | 1546.87 |
| | 6 | –611.62 | 79 | 1381.25 | 1588.62 |
| STFA | 1 | –1186.39 | 35 | 2442.77 | 2534.65 |
| | 2 | –1062.48 | 46 | 2216.95 | 2337.70 |
| | 3 | –689.53 | 56 | 1491.05 | 1638.05 |
| | **4** | **–564.75** | **65** | **1259.50** | **1430.12** |
| | 5 | –562.26 | 73 | 1270.52 | 1462.14 |
| | 6 | –562.21 | 80 | 1284.43 | 1494.42 |

Table 3: Summary ML results together with the associated standard errors in parentheses for the best chosen model

| Variable | Parameter | | | | | |
|---|---|---|---|---|---|---|
| | $\boldsymbol{\mu}$ | $\boldsymbol{B}$ | | | | $\boldsymbol{d}$ |
| rcc | –0.035 | –0.039 | [0.573] | 0.371 | –0.066 | 0.2074 |
| | (0.091) | (0.087) | (0.080) | (0.094) | (0.080) | (0.0319) |
| wcc | –0.015 | –0.024 | –0.062 | 0.360 | –0.019 | 0.7132 |
| | (0.099) | (0.087) | (0.096) | (0.098) | (0.100) | (0.1179) |
| Hc | –0.031 | –0.031 | [0.682] | 0.402 | 0.017 | 0.0164 |
| | (0.089) | (0.083) | (0.077) | (0.091) | (0.076) | (0.0173) |
| Hg | –0.040 | –0.020 | [0.640] | 0.393 | 0.126 | 0.1268 |
| | (0.089) | (0.084) | (0.078) | (0.095) | (0.081) | (0.0252) |
| Fe | –0.039 | 0.004 | –0.122 | 0.162 | 0.205 | 0.7410 |
| | (0.100) | (0.093) | (0.094) | (0.098) | (0.098) | (0.1171 ) |
| bmi | –0.019 | [0.501] | –0.038 | [0.445] | 0.384 | 0.0025 |
| | (0.064) | (0.056) | (0.056) | (0.072) | (0.050) | (0.0011) |
| ssf | 0.020 | 0.070 | –0.278 | [0.689] | 0.029 | 0.0410 |
| | (0.072) | (0.055) | (0.062) | (0.058) | (0.057) | (0.0068) |
| Bfat | 0.028 | 0.017 | –0.296 | [0.706] | 0.074 | 0.0023 |
| | (0.074) | (0.054) | (0.061) | (0.055) | (0.058) | (0.0024) |
| lbm | –0.005 | [0.825] | –0.041 | 0.304 | –0.113 | 0.0012 |
| | (0.043) | (0.042) | (0.059) | (0.086) | (0.033) | (0.0003) |
| Ht | 0.033 | [0.620] | –0.134 | 0.192 | –0.673 | 0.0001 |
| | (0.052) | (0.047) | (0.062) | (0.089) | (0.040) | (0.0019) |
| Wt | -0.002 | [0.723] | –0.103 | 0.429 | –0.074 | 0.0002 |
| | (0.041) | (0.041) | (0.055) | (0.079) | (0.030) | (0.0001) |
| | | $\boldsymbol{\lambda}$ | | | | $\nu$ |
| | | –0.874 | –4.042 | 6.531 | 1.283 | 6.28 |
| | | (0.468) | (1.437) | (1.808) | (0.366) | (1.620) |

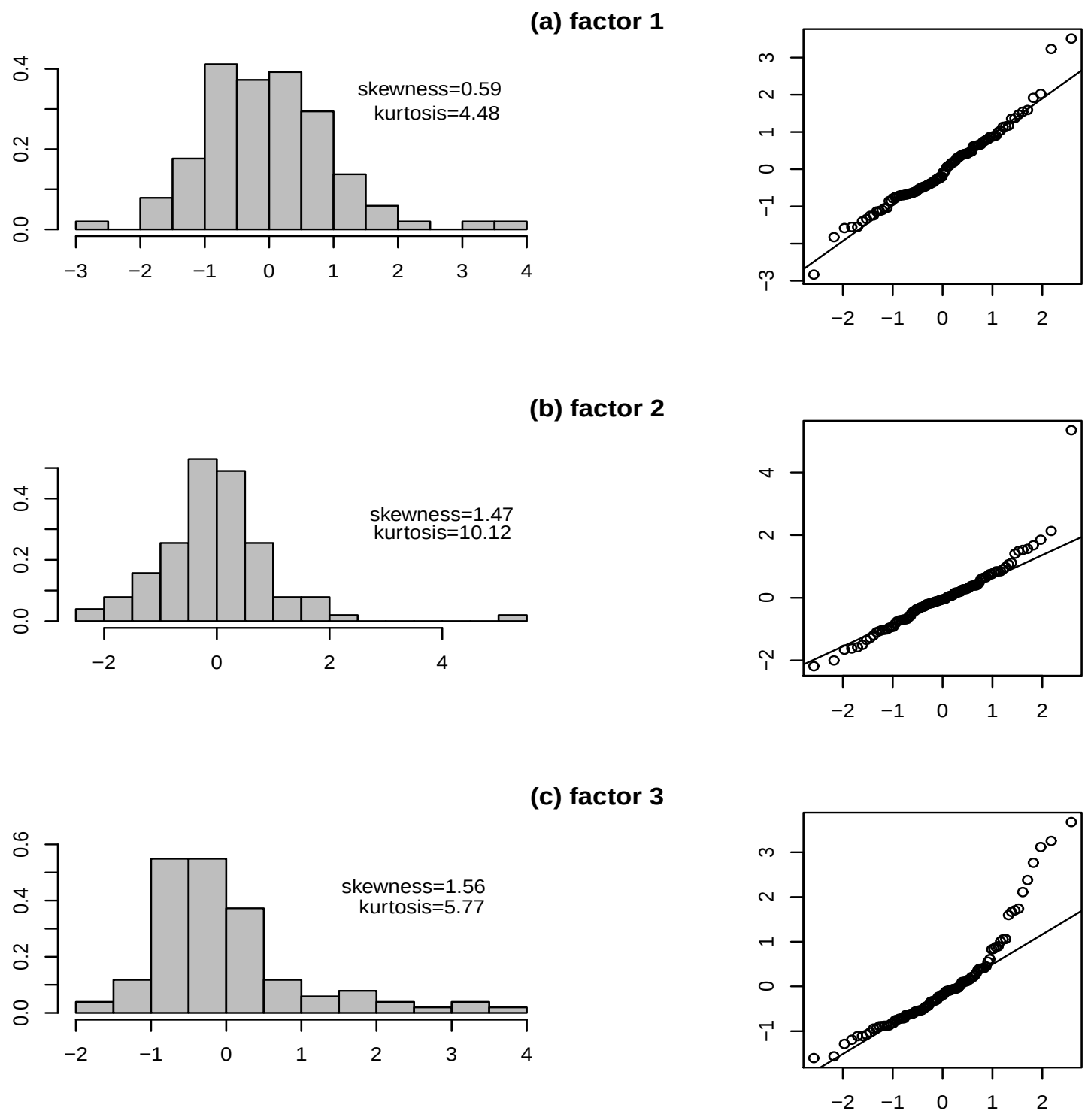

Figure 1: Histograms and corresponding normal quantile plots of the estimated factor scores obtained from fitting FA to $n = 102$ male athletes of AIS data.

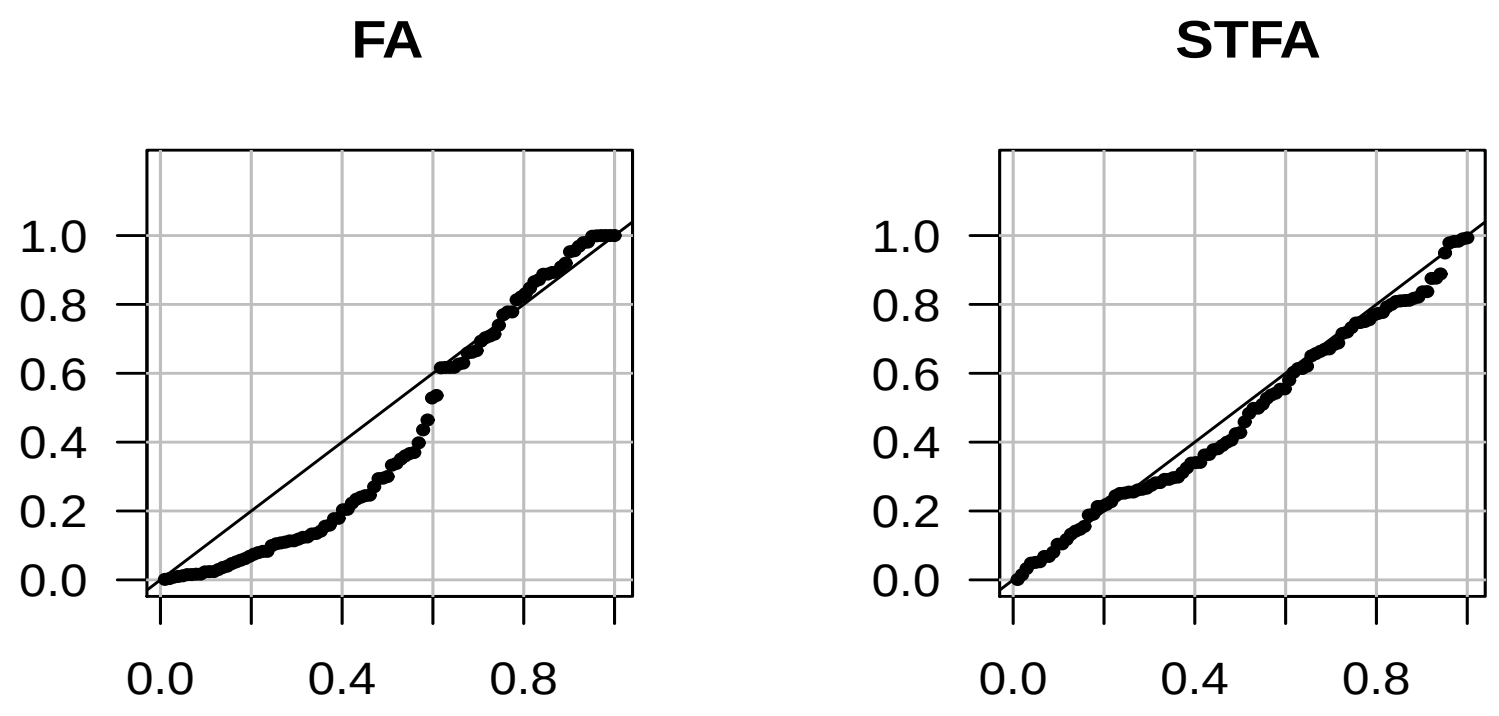

Figure 2: Healy's plot for assessing the goodness-of-fit of fitted models.

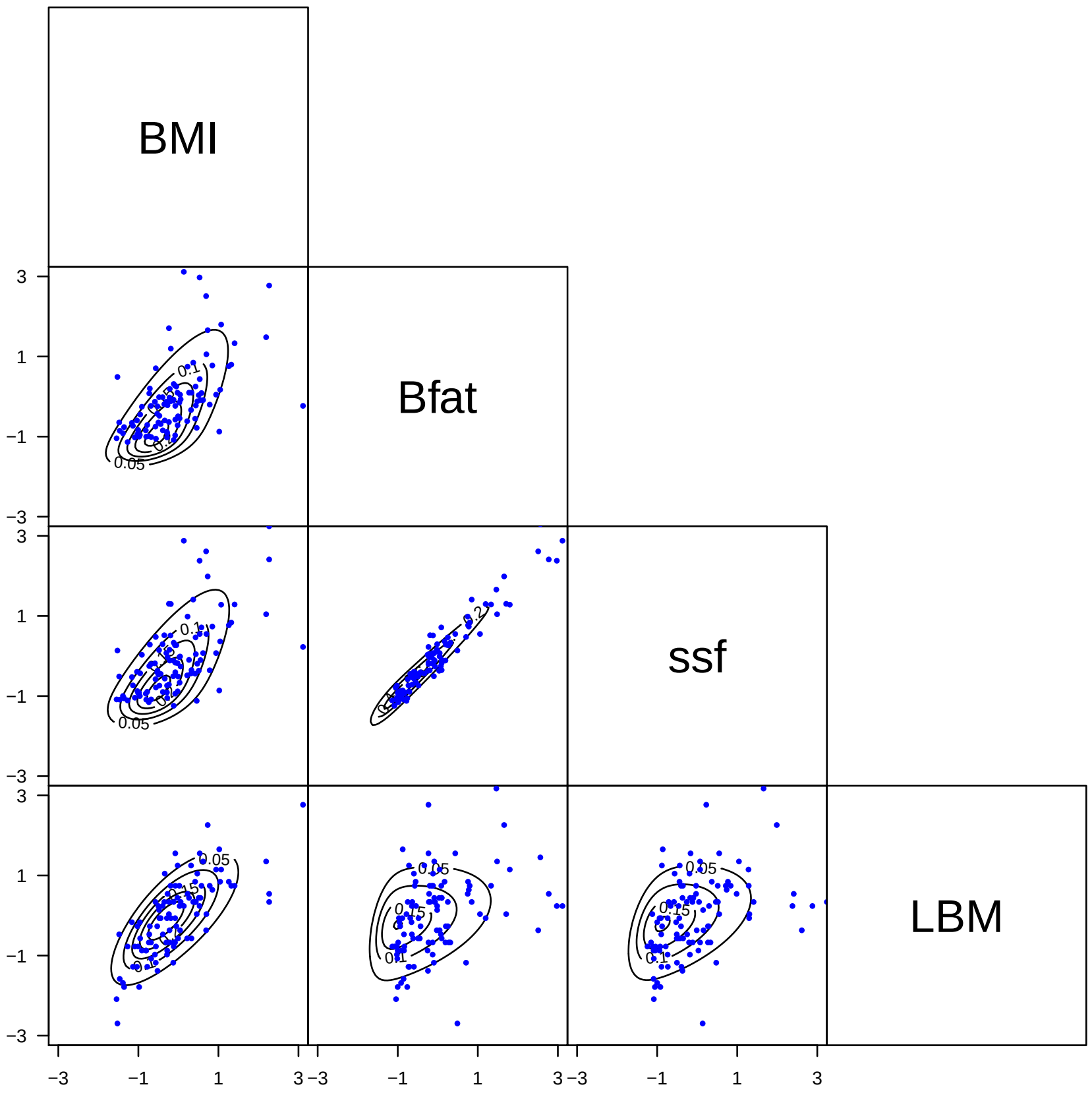

Figure 3: Scatter plots of pairs of four selected variables of 102 male AIS athletes and coordinate projected contours.